\documentclass[10pt,journal,compsoc]{IEEEtran}
\usepackage{algorithm}
\usepackage{algorithmic}
\usepackage{subcaption}
\usepackage{multirow}
\usepackage{multicol}
\usepackage{tabularx}
\usepackage{float}

\newcolumntype{Y}{>{\centering\arraybackslash}X}

\usepackage{amsfonts}
\usepackage{amsthm}
\usepackage{amsmath}
\usepackage{caption}
\usepackage{array}
\usepackage{booktabs}
\usepackage{makecell}
\usepackage{graphicx}
\theoremstyle{definition}
\newtheorem{definition}{Definition}[section]

\theoremstyle{theorem}
\newtheorem{theorem}{Theorem}[section]
\newtheorem{remark}{Remark}

\usepackage{pgfplots}

\ifCLASSOPTIONcompsoc
  \usepackage[nocompress]{cite}
\else
  \usepackage{cite}
\fi
\ifCLASSINFOpdf
\else
\fi
\begin{document}
%
\title{Edgeless-GNN: Unsupervised Representation Learning for Edgeless Nodes}
%
%
%
%

\author{Yong-Min Shin, {\em Student Member}, {\em IEEE}, Cong Tran, {\em Member}, {\em IEEE}, \\Won-Yong Shin, {\em Senior Member}, {\em IEEE},  and  Xin Cao
\IEEEcompsocitemizethanks{\IEEEcompsocthanksitem Y.-M. Shin and W.-Y. Shin are with the School of Mathematics and Computing (Computational Science and Engineering), Yonsei University, Seoul 03722, Republic of Korea. \protect\\
E-mail: \{jordan3414, wy.shin\}@yonsei.ac.kr.
\IEEEcompsocthanksitem C. Tran was with the Department of Computer Science and Engineering, Dankook University, Yongin 16890, Republic of Korea, and also with the Machine Intelligence \& Data Science Laboratory, Yonsei University, Seoul 03722, Republic of Korea. He is now with the Faculty of Information Technology, Posts and Telecommunications Institute of Technology, Hanoi 100000, Vietnam. \protect\\
E-mail: congtt@ptit.edu.vn
\IEEEcompsocthanksitem X. Cao is with the School of Computer Science and Engineering, The University of New South Wales, Sydney 2052, Australia.\protect\\
E-mail:  xin.cao@unsw.edu.au.\protect\\
(Corresponding author: Won-Yong Shin.)}}
%
%

\markboth{}%
{Shell \MakeLowercase{\textit{et al.}}: Edgeless-GNN: Unsupervised Inductive Edgeless Network Embedding}
%



\IEEEtitleabstractindextext{%
\begin{abstract}
We study the problem of embedding {\em edgeless} nodes such as users who newly enter the underlying network, while using graph neural networks (GNNs) widely studied for effective representation learning of graphs. Our study is motivated by the fact that GNNs cannot be straightforwardly adopted for our problem since message passing to such edgeless nodes having no connections is impossible. To tackle this challenge, we propose \textsf{Edgeless-GNN}, a novel inductive framework that enables GNNs to generate node embeddings even for edgeless nodes through {\em unsupervised learning}. Specifically, we start by constructing a proxy graph based on the similarity of node attributes as the GNN's computation graph defined by the underlying network. The known network structure is used to train model parameters, whereas a {\em topology-aware} loss function is established in such a way that our model judiciously learns the network structure by encoding positive, negative, and second-order relations between nodes. For the edgeless nodes, we {\em inductively} infer embeddings by expanding the computation graph. By evaluating the performance of various downstream machine learning tasks, we empirically demonstrate that \textsf{Edgeless-GNN} exhibits (a) superiority over state-of-the-art inductive network embedding methods for edgeless nodes, (b) effectiveness of our topology-aware loss function, (c) robustness to incomplete node attributes, and (d) a linear scaling with the graph size.
\end{abstract}

}

\maketitle

\IEEEdisplaynontitleabstractindextext

%
\IEEEpeerreviewmaketitle

\IEEEraisesectionheading{\section{Introduction}\label{sec:introduction}}

%
%
%
%
\subsection{Background and Motivation}
\IEEEPARstart{G}{raphs} are a ubiquitous way to organize a diverse set of real-world data such as social networks, citation networks, molecular graph structures, and recommender systems.\footnote{In the following, we use the terms "graph" and "network" interchangeably.} Moreover, nodes in graphs are often associated with rich attribute information~\cite{qi2012attnet}, which motivates researchers to leverage both topological and attribute information to solve various tasks. In recent years, graph neural networks (GNNs)~\cite{scarselli2009gnn,gilmer2017neural,DBLP:conf/iclr/KipfW17,DBLP:conf/nips/HamiltonYL17,velivckovic2017graph,xu2018powerful,wu2019survey} have been widely studied as a powerful means to extract useful low-dimensional features from attributed graphs while performing various downstream graph mining tasks such as node classification~\cite{DBLP:conf/iclr/KipfW17,shi2020mlne,shen2021nt}, link prediction~\cite{zhang2018seal,you2019pgnn}, and community detection~\cite{bianchi2020mincutpool, he2020mafasgcn}. GNNs have become a successful graph representation learning model due to their high expressive capability via message passing~\cite{gilmer2017neural}, which exchanges latent information of nodes through edges for acquiring richer representations.

Nevertheless, relatively little attention has been devoted to discovering embeddings of {\em edgeless nodes} ({\em i.e., structure-unaware nodes}) whose topological information is not available. This lack of topological information may occur in several real-world networks extracted from various systems. There are several sources of incompleteness for the network structure in real-world scenarios. First, in co-authorship networks, some papers may be contributed by single (or isolated) authors, creating nodes without any connections to others. For example, a statistical analysis showed that 15.48\% of the published papers in the field of information retrieval during the period of 2001--2008 were single-authored papers~\cite{singleauthor}. Second, in other real-world networks including social networks, some users may completely hide their friendships due to privacy settings specified by such users~\cite{dey2012facebook, buccafurri2015facebook}. As an example, a demographic analysis of Facebook users in New York City in June 2011 demonstrated that 52.6\% of the users hid the lists of Facebook friends~\cite{dey2012facebook}. Despite the absence of connectivity information associated with such hidden or new nodes, it would be still possible to perform downstream graph mining tasks when the attribute information of nodes ({\it e.g.}, authors' biographical features and research interests) is available since such information is much easier to acquire through publicly available sources such as Google Scholar. 

However, simply adopting GNN models to learn the representations of such edgeless nodes is not straightforward and poses two major challenges. First and foremost, edgeless nodes literally have no edges whereas edges are essential to perform message passing. Although existing GNNs may technically operate on edgeless nodes by omitting message passing from/to such nodes due to the topology unawareness, this will cause GNNs to loose their expressive power. Second, an appropriate loss function needs to be designed to precisely discover representations for edgeless nodes by training the underlying GNN model, which is not straightforward especially for unsupervised learning settings. Even if the analysis of networks harnessing node attributes has emerged as another research area, little attention has been paid to the {\it inductive} representation learning technique for more challenging yet practical situations in which an underlying attributed graph is incomplete with edgeless nodes. Although prior studies such as Graph2Gauss~\cite{bojchevski2017deep} and DEAL~\cite{hao2020inductive} attempted to solve the aforementioned problem in the context of inductive link prediction, they adopted multilayer perceptron (MLP) encoders as the base architecture and the performance of downstream tasks other than link prediction was unexplored.

\subsection{Main Contributions}
In this paper, we consider a practical scenario of {\it attributed} networks in which a portion of nodes have no available edges, {\it i.e.}, topological information. In this attributed network model, we study the problem of embedding such {\it edgeless} nodes ({\it e.g.}, users who newly enter the underlying network) through GNNs. More specifically, we are interested in {\it inductively} and {\it unsupervisedly} discovering vector representations of edgeless nodes by effectively designing an entirely new GNN framework along with a new loss function for model optimization.

\begin{figure}[!t]
\centering
\includegraphics[width=\columnwidth]{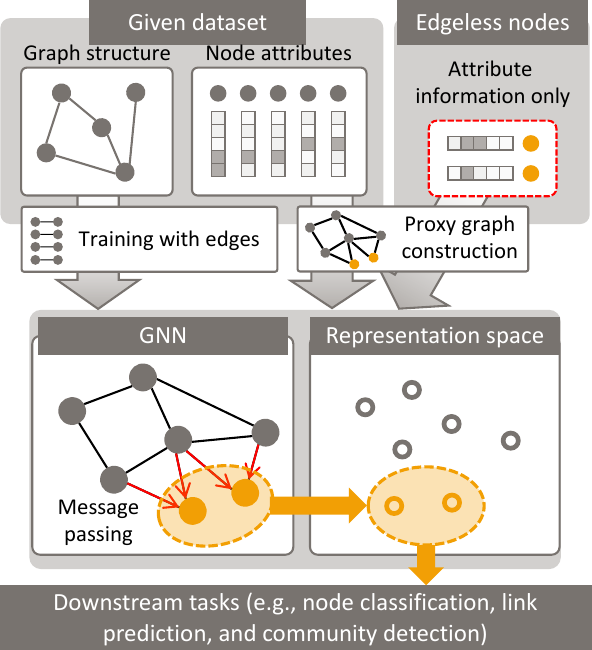}
\caption{A brief sketch of our \textsf{Edgeless-GNN} framework.}
\label{firstfigure}
\end{figure}

Motivated by the wide applications of GNNs to attributed networks and their generalization abilities, a natural question arising is: "Will existing GNN models be indeed applicable and beneficial for solving the problem of inductive embedding for edgeless nodes?" To answer this question, we present \textsf{Edgeless-GNN}, a novel framework that enables GNNs to generate vector representations for edgeless nodes in attributed networks, which is used for conducting various downstream tasks. Fig.~\ref{firstfigure} illustrates a brief sketch of our \textsf{Edgeless-GNN} framework. First, we construct a proxy graph, which is built upon the similarity between node attributes as a means of replacing the GNN's original computation graph defined by the given network structure. We note that this new proxy graph is generated under the {\it homophily} assumption that adjacent nodes in the underlying network tend to have similar node attributes. Second, more importantly, we propose a {\em topology-aware} loss function for {\it unsupervisedly} training model parameters without any node labels, which makes full use of the known network structure. In other words, instead of the attribute-aided proxy graph, we fully exploit the underlying network structure for model training. Our topology-aware loss function is inspired by the mechanism of contrastive learning over graphs in a sophisticated manner. Specifically, our new loss function exploits not only the first-order proximity of node pairs but also second-order positive relations in learning node embeddings. This allows the model to learn the topological structure of the given network while performing message passing over newly created edges in the proxy graph. Since our \textsf{Edgeless-GNN} framework is inductive, it is possible to directly infer representations for edgeless nodes by expanding the computation graph connecting the nodes with edges and the edgeless nodes using node attributes in the same manner to facilitate the feed-forward process of GNNs. Additionally, we analyze the computational complexity of the \textsf{Edgeless-GNN} framework, which is shown to scale linearly with the network size. Since our \textsf{Edgeless-GNN} does not assume a specific GNN architecture, various GNNs in the literature~\cite{DBLP:conf/iclr/KipfW17, DBLP:conf/nips/HamiltonYL17, xu2018powerful} can be adopted in a plug-and-play fashion. This implies that our framework is GNN-model-agnostic; thus, GNN models can be appropriately chosen in our \textsf{Edgeless-GNN} framework according to one's needs and graph mining tasks. Moreover, in contrast to the edgeless nodes, in case of structure-aware nodes, we perform downstream tasks by appropriately choosing an existing GNN model since message passing is possible alongside the observable edges.

To validate the superiority and effectiveness of our \textsf{Edgeless-GNN} framework, we comprehensively perform empirical evaluations for various real-world benchmark datasets. First, experimental results show that our framework consistently outperforms state-of-the-art inductive network embedding approaches of edgeless nodes for almost all cases when we carry out three downstream tasks such as link prediction, node classification, and community detection. Second, interestingly, it is observed that simply adopting the existing GNN's loss function while using a proxy graph as the computation graph indeed fails to guarantee satisfactory performance and is even far inferior to a na\"ive baseline method employing node attributes only. This clearly justifies the need of a new GNN framework for edgeless nodes. This also implies that our newly established loss function plays a very crucial role to successfully infer embeddings of the edgeless nodes by bridging the structural and attribute information. Third, our experimental results demonstrate the robustness of our \textsf{Edgeless-GNN} framework to a more difficult and challenging situation where a large portion of node attributes are missing. Fourth, we investigate the impact of hyperparameters by confirming that both the second-order proximity loss term and the parameter controlling negative node pairs in our loss are vital for the training model to infer high-quality vector representations. Finally, we empirically validate our complexity analysis.

The main contributions of this paper are summarized as follows:
\begin{itemize}
    \item We propose \textsf{Edgeless-GNN}, a novel GNN-agnostic representation learning framework for networks with edgeless nodes.
    \item As core components of \textsf{Edgeless-GNN}, we introduce not only the inductive-learning-enabled construction of a proxy graph but also the design of a topology-aware loss function for unsupervised learning.
    \item We comprehensively validate the superiority and effectiveness of our proposed \textsf{Edgeless-GNN} framework through extensive experiments using five real-world attributed networks.
    \item We analyze and empirically show the computational complexity of \textsf{Edgeless-GNN}.
\end{itemize}
Our methodology sheds light on how to effectively discover vector representations even when no topological structure of some nodes is available. 

\begin{table}
  \caption{Summary of notations.}
  \label{NotationTable}
  \scalebox{1}{\begin{tabularx}{\columnwidth}{ll}
        \toprule
        \textbf{Notation} & \textbf{Description} \\
        \midrule
        $G$ & Given attributed network\\
        \rule{0pt}{10pt}$\mathcal{V}$ & Set of nodes in $G$\\
        \rule{0pt}{10pt}$\mathcal{V}'$ & Set of edgeless nodes\\
        \rule{0pt}{10pt}$\mathcal{V}^{\textnormal{all}}$ & $\mathcal{V} \cup \mathcal{V}'$\\
        \rule{0pt}{10pt}$\mathcal{E}$ & Set of edges in $G$\\
        \rule{0pt}{10pt}$\mathcal{X}$ & Set of attributes of nodes in $\mathcal{V}$\\
        \rule{0pt}{10pt}$\mathcal{X}'$ & Set of attributes of nodes in $\mathcal{V}'$\\
        \rule{0pt}{10pt}$\mathcal{X}^{\textnormal{all}}$ & $\mathcal{X} \cup \mathcal{X}'$\\
        \rule{0pt}{10pt}$G_{p}$ & Proxy graph constructed from $\mathcal{X}$\\
        \rule{0pt}{10pt}$G_{p}^{\textnormal{all}}$ & Proxy graph constructed from $\mathcal{X}^{\textnormal{all}}$\\
        \rule{0pt}{10pt}$\mathcal{E}_{p}$ & Set of edges from $G_{p}$\\
        \rule{0pt}{10pt}$\mathcal{E}_{p}^{\textnormal{all}}$ & Set of edges from $G_{p}^{\textnormal{all}}$\\
        \rule{0pt}{10pt}$\mathbf{Z}$ & Embedding for the nodes in $\mathcal{V}$\\
        \rule{0pt}{10pt}$\mathbf{Z}'$ & Embedding for the nodes in $\mathcal{V}'$\\
        \bottomrule
\end{tabularx}}
\end{table}

\section{Related work} \label{relatedwork}
The framework that we propose in this study is related to two broader topics of research, namely inductive network embedding and GNNs.

\textbf{Inductive network embedding.} Inductive network embedding (more specifically, learning node representations) has been widely studied before the popularity of GNNs. Planetoid~\cite{DBLP:conf/icml/YangCS16} proposed a model, named Planetoid-I, which was designed for generating node embedding vectors in the inductive setting. Graph2Gauss~\cite{bojchevski2017deep} adopted an MLP architecture while using a rank-based loss. More recently, DEAL~\cite{hao2020inductive} was modeled by constructing both attribute-oriented and structure-oriented encoders and then aligning two types of embeddings via two encoders. We note that, in~\cite{bojchevski2017deep,hao2020inductive}, the problem of {\em inductive link prediction} was addressed when the network structure of new nodes is unknown.

\textbf{GNNs.} The early model of GNNs was first proposed by~\cite{scarselli2009gnn} using a recurrent neural network model. More recently, GCN~\cite{DBLP:conf/iclr/KipfW17} proposed an efficient way to learn convolutional filters on graphs. In GraphSAGE~\cite{DBLP:conf/nips/HamiltonYL17}, various aggregation methods such as average, max pooling, and long short-term memory (LSTM) were proposed along with neighborhood sampling. GAT~\cite{velivckovic2017graph} was presented by synthesizing self-attention layers, where different importances are assigned to neighboring nodes. By analyzing the expressive capability of popular GNN models, an architecture exhibiting a further representational power was designed in~\cite{xu2018powerful}. JK-Nets~\cite{xu2018representation} was designed by using intermediate layer-aggregation mechanisms for better representation learning. DGL~\cite{velickovic2019dgl} utilized a mutual information maximization approach for training GNNs. Several studies were also carried out to take advantage of graph construction based on attribute information alongside GNNs. AM-GCN~\cite{wang2020gcn} proposed a multi-channel GCN model by learning node representations based on not only the network structure but also the $k$-nearest-neighbor ($k$NN) graph generated from node attributes. In \cite{franceschi2019lds}, the unavailable network structure and the parameters of the GCN model were jointly learned by solving a bilevel programming problem. Recent attempts include not only efficient designs of GNN models in~\cite{chen2018fastgcn,zang2020graphsaint} but also designs of very deep GCNs for a further performance boost in~\cite{li2019deepgcn, chiang2019clustergcn}. Furthermore, GLNN~\cite{zhang2022graphless} developed an MLP model to be trained by a teacher GNN model to boost the performance.

\textbf{Discussion.}
Despite these contributions, it has been largely underexplored in the literature how to exploit the power of GNNs in the context of representation learning for {\em edgeless} nodes. Although GNNs are built upon message passing mechanisms that are shown to flexibly model complex interactions among nodes, they cannot be straightforwardly employed to solve our problem since no information can be passed from/to the edgeless nodes. Moreover, recent studies on \textit{inductive link prediction} in \cite{bojchevski2017deep,hao2020inductive} are limited only to MLP architectures and thus do not take advantage of potentials of rather powerful GNNs. Compared to our study, applications to other downstream tasks such as node classification and community detection were not studied in~\cite{hao2020inductive}. Node classification problems were shown in~\cite{bojchevski2017deep} only for transductive learning settings.

\begin{figure*}[!t]
\centering
\includegraphics[width=\textwidth]{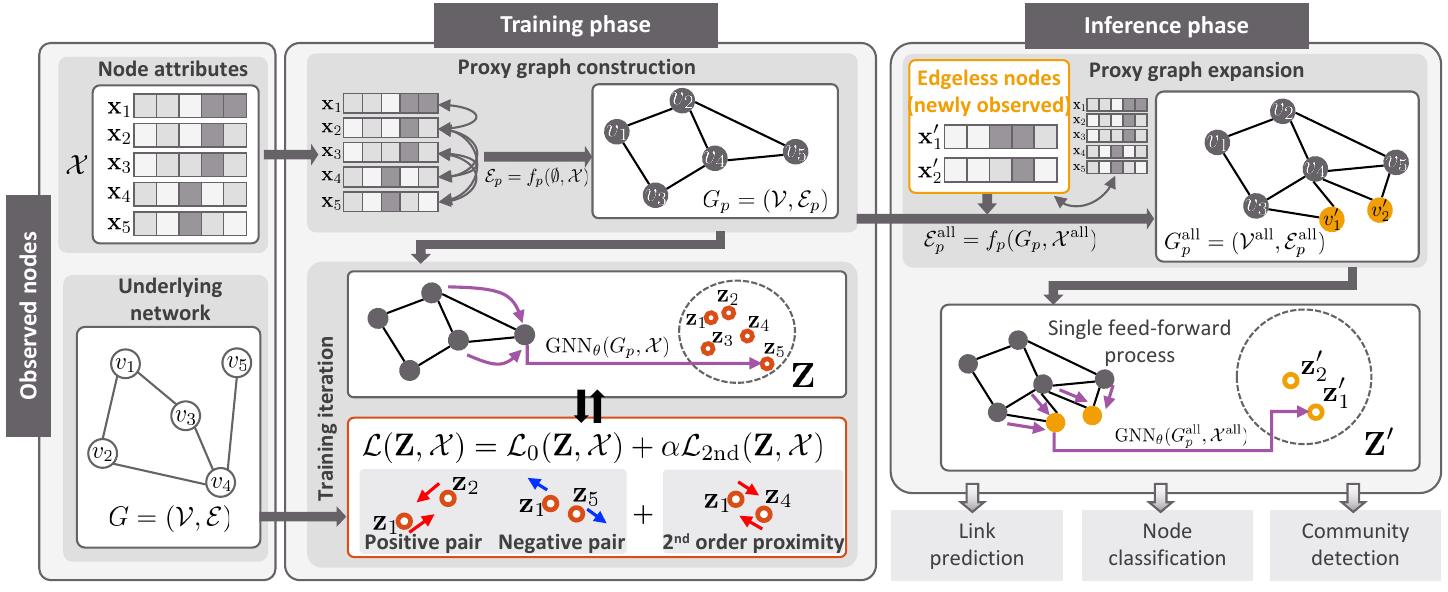}
\caption{A schematic overview of our \textsf{Edgeless-GNN} framework.}
\label{Schematicoverview}
\end{figure*}

\section{Preliminaries} \label{prelimaries}
In this section, we describe our basic setting along with the notations used in the paper, followed by the formal definition of the problem. Then, we describe the architecture of GNNs in general.

\subsection{Basic settings}
Let us denote a given network as $G = (\mathcal{V},\mathcal{E})$, where $\mathcal{V}$ is set of $N$ nodes and $\mathcal{E}$ is the set of edges between pairs of nodes in $\mathcal{V}$. We assume $G$ to be an undirected unweighted {\em attributed} network without self-loops or repeated edges. We define $\mathbf{x}_{i} \in \mathbb{R}^{f}$ as the attribute vector of node $v_{i} \in \mathcal{V}$, and $\mathcal{X} = \{\mathbf{x}_{1}, \cdots,\mathbf{x}_{N}\}$ as the set of node attribute vectors, where $f$ is the number of attributes per node.

In the inductive learning setting, we would like to newly introduce a set of $M$ edgeless nodes, denoted as $\mathcal{V}'$, which has not been seen yet during the training phase.\footnote{In our study, although the existence of edgeless nodes is known beforehand, we treat them as newly given since existing GNN models cannot provide a way of integrating the edgeless nodes into the underlying network.} Each of these $M$ nodes has an associated attribute vector $\mathbf{x}'_{i} \in \mathbb{R}^{f}$ for $i\in\{1,\cdots,M\}$. We denote the set of attributes of these $M$ nodes as $\mathcal{X}' = \{\mathbf{x}'_{1}, \cdots,\mathbf{x}'_{M}\}$. However, the network structure of these $M$ nodes in $\mathcal{V}'$ is unavailable, which is a feasible scenario ({\em e.g.}, a co-authorship network in which some nodes are single-authored papers). In other words, two types of edges, including the edges connecting two nodes in $\mathcal{V}'$ and the edges connecting one node in $\mathcal{V}$ and another node in $\mathcal{V}'$, are not given beforehand, as in \cite{hao2020inductive}. In this context, these nodes in $\mathcal{V}'$ are the so-called {\em edgeless} nodes. The inductive setting is interested in learning representations of $\mathcal{V}'$ while the observed nodes in $\mathcal{V}$ are not of interest since their representations can be learned by straightforwardly adopting existing GNN models.

\subsection{Problem definition}
\theoremstyle{definition}
\begin{definition}[\textbf{\textbf{Inductive embedding of edgeless nodes}}]
    Given a network $G = (\mathcal{V},\mathcal{E})$ and two sets of node attributes, $\mathcal{X}$ and $\mathcal{X}'$, inductive embedding of {\em edgeless} nodes aims to discover vector representations of {\em edgeless} nodes in such a way that the embedding vectors unsupervisedly encode the unseen structural information and the available attribute information.
\end{definition}

\subsection{GNN Architecture}
GNNs operate on a computation graph $G_c$, which determines the flow of information in the message passing mechanism~\cite{gilmer2017neural,DBLP:conf/nips/HamiltonYL17,xu2018powerful,xu2018representation}. The typical choice of $G_{c}$ is the underlying network itself, {\em i.e.}, $G_{c} = G$. In each layer, GNN models update the representation of a node by aggregating latent representations of its neighbors using two functions with learnable parameters, namely AGGREGATE and UPDATE. Formally, at the $p$-th layer of a GNN, $\textnormal{AGGREGATE}^{(p)}$ aggregates (latent) feature information from the local neighborhood of node $v_i$ in the computation graph $G_c$ as follows:
\begin{equation}
\label{aggregateequation}
    \mathbf{m}^{p}_{i} \leftarrow \textnormal{AGGREGATE}^{(p)}(\{\mathbf{h}_{j}^{p-1}|v_j \in \mathcal{N}_{i} \cup \{v_i\} \}),
\end{equation}
where ${\bf h}^{p-1}_j$ denotes the latent representation vector of node $v_j$ at the $(p-1)$-th layer, $\mathcal{N}_{i}$ indicates the set of neighbor nodes of $v_{i}$ in $G_{c}$, and ${\bf m}_i^p$ is the aggregated information at the $p$-th layer. We note that self-loops are typically added to the computation graph $G_{c}$ for self-information preservation. In the update step, the latent representation at the next layer is produced by using each node and its aggregated information from $\textnormal{AGGREGATE}^{(p)}$ as follows:
\begin{equation} \label{updateequation}
    \mathbf{h}^{p}_{i} \leftarrow \textnormal{UPDATE}^{(p)}(\mathbf{h}_{i}^{p-1},\mathbf{m}^{p}_{i}).
\end{equation}
Additionally, for each node $v_i$, the node attributes ${\bf x}_i \in \mathcal{X}$ are initially used as the representation vector ({\em i.e.}, $\mathbf{h}^{0}_{i} = \mathbf{x}_{i}$), and the representation at the final layer is the embedding vector ${\bf z}_i$. In our paper, ${\bf Z}$ denotes the embedding matrix whose $i$-th row corresponds to ${\bf z}_i$.

\begin{remark}
Now, let us state how the above two functions AGGREGATE and UPDATE in (\ref{aggregateequation}) and (\ref{updateequation}), respectively, can be specified by several types of GNN models. As one of the commonly used GNN models, GCN~\cite{DBLP:conf/iclr/KipfW17} can be implemented by using
\begin{align}
    \textnormal{AGGREGATE}^{(p)}_{i} &= \sum_{j} \dfrac{1}{\sqrt{\textnormal{degree}(i)+1}\sqrt{\textnormal{degree}(j)+1}}\mathbf{h}^{p-1}_{j}\\
    \textnormal{UPDATE}^{(p)}_{i} &= \sigma(\mathbf{W}^{p}\cdot\mathbf{m}^{p}_{i}),
\end{align}
where $\textnormal{degree}(\cdot)$ indicates the degree of each node and $\sigma(\cdot)$ is the activation function.
In addition, as one of powerful GNN models, GraphSAGE~\cite{DBLP:conf/nips/HamiltonYL17} with the mean aggregator can be designed by setting
\begin{align}
    \textnormal{AGGREGATE}^{(p)}_{i} &= \dfrac{1}{\textnormal{degree}(i)+1}\sum_{j}\mathbf{h}^{p-1}_{j}\\
    \label{graphsageupdate}
    \textnormal{UPDATE}^{(p)}_{i} &= \sigma(\mathbf{W}^{p}\cdot \textnormal{concat}(\mathbf{h}^{p-1}_{i}, \mathbf{m}^{p}_{i})),
\end{align}
where $\textnormal{concat}(\cdot,\cdot)$ is the concatenation operation of two input vectors and $\mathbf{W}^{p}$ is a learnable weight matrix. Other popular GNN variants such as GAT~\cite{velivckovic2017graph} and GIN~\cite{xu2018powerful} can also be specified according to their designed function settings. Note that, for GCN and GraphSAGE with $L$ layers, the model parameters $\theta$ can be expressed as the set of weights $\{\mathbf{W}^{p}\}_{p=1,\cdots,L}$.
\end{remark}

\section{Methodology} \label{methodology}
In this section, we describe the proposed \textsf{Edgeless-GNN} framework including the design of our own loss function to solve the problem of inductive embedding for edgeless nodes. The schematic overview of \textsf{Edgeless-GNN} is illustrated in Fig.~\ref{Schematicoverview}.

\subsection{\textsf{Edgeless-GNN} Framework}

In this subsection, we explain our \textsf{Edgeless-GNN} framework along with the proxy graph construction, which is a core component of our framework.

In the feed-forward process of GNNs, the typical choice of $G_{c}$ is the underlying network itself, {\em i.e.}, $G_{c} = G$. This approach cannot be adopted as $G$ does not have any connections to edgeless nodes $\mathcal{V}'$. To overcome this problem, we construct two proxy graphs $G_p = (\mathcal{V},\mathcal{E}_{p})$ and $G_p^{\textnormal{all}} = (\mathcal{V}^\textnormal{all},\mathcal{E}^{\textnormal{all}}_{p})$ as an alternative to the GNN's original computation graph $G_c$ (refer to line 1 in Algorithm \ref{mainalgorithm}). Edges of the proxy graphs, $\mathcal{E}_p$ and $\mathcal{E}_p^{\textnormal{all}}$, are created by the same proxy graph construction process, which we can describe as a function with two inputs $f_p(\cdot, \cdot)$. The input of $f_p(\cdot, \cdot)$ includes a computation graph to be expanded ($\emptyset$ if we build one from scratch) and a set of node attributes, respectively. Formally, we use $f_p(\cdot, \cdot)$ to create $\mathcal{E}_{p} = f_p(\emptyset, \mathcal{X})$ and $\mathcal{E}^{\textnormal{all}}_{p} = f_p(G_p, \mathcal{X}^{\textnormal{all}})$. In our framework, the resulting proxy graphs are used as the computation graph, {\em i.e.}, $G_c = G_p$ during training and $G_c = G_p^{\text{all}}$ during inference. 

The key idea behind the proxy graph construction is to 1) use node attributes as the main ingredient for proxy graphs such that edgeless nodes can also be fed into the feed-forward process of GNNs and 2) use the same function $f_p(\cdot, \cdot)$ for both $G_p$ and $G_p^{\textnormal{all}}$ such that the GNN model will consistently harness a computation graph and its expansion during training and inference, respectively. Note that, designing a rather sophisticated learning module to find the set of appropriate edges to connect edgeless nodes corresponds to basically solving the link prediction task, which should be preceded by discovering vector representations via GNNs. Thus, we instead construct a proxy graph where the GNN model acts upon.

During the training phase, the proxy graph $G_p$ enables us to acquire representations for $\mathcal{V}$ as follows:
\begin{equation}
\label{trainembspace}
    \mathbf{Z} = \textsf{GNN}_{\theta}(G_p, \mathcal{X}, \mathcal{V}),
\end{equation}
where the third argument of $\textsf{GNN}_\theta (\cdot,\cdot,\cdot)$ represents the set of nodes of interest whose representations are to be calculated. Here, each row of ${\bf Z}\in \mathbb{R}^{N\times d}$ indicates the vector representation of each node in $\mathcal{V}$; $d$ is the dimension of the representation space; and $\theta$ represents the set of model parameters of the GNN model used in \textsf{Edgeless-GNN}. Now, $\theta$ is learned by our loss function $\mathcal{L}({\bf Z}, \mathcal{E})$, which will be specified in Section~\ref{subsection:loss}. In other words, the embedding ${\bf Z}$ is first found by running a GNN model on the constructed $G_{p}$ from the node attributes and is then optimized by leveraging the local network structure of each node ({\em i.e.}, multi-hop neighbors as well as direct neighbors) in the underlying network $G$. In this fashion, the model learns how to incorporate the topological information of $G$ while using computation graphs generated from $f_p(\cdot, \cdot)$.

Next, we turn to the inference phase, which inductively finds representations for the edgeless nodes in $\mathcal{V}'$. The set of node attributes including edgeless nodes, $\mathcal{X}^{\textnormal{all}} = \mathcal{X} \cup \mathcal{X}'$, are fed to the trained GNN model to calculate representations ${\bf Z}'\in \mathbb{R}^{M\times d}$ for the edgeless nodes $\mathcal{V}'$ by a simple feed-forward computation as follows:
\begin{equation}
\label{testembspace}
    \mathbf{Z}' = \textsf{GNN}_{\theta}(G_{p}^{\textnormal{all}}, \mathcal{X}^{\textnormal{all}}, \mathcal{V}').
\end{equation}
Representations of the nodes in $\mathcal{V}'$ can be efficiently calculated by taking into account the nodes of interest required only for message passing~\cite{DBLP:conf/nips/HamiltonYL17}. Finally, we are able to perform various downstream tasks using $\mathbf{Z}'$ ({\em i.e.}, the embedding vectors of the edgeless nodes in $\mathcal{V}'$). 

In our implementation, we choose a $k$NN graph ($k$NNG) as the proxy graph construction function $f_p(\cdot, \cdot)$, which is most straightforward among graph construction strategies. Additionally, for $k$NNG construction, we use the cosine similarity to measure the similarity between two node attributes. 

\subsection{Model Training with Topology-Aware Loss} \label{subsection:loss}

\begin{algorithm}[t]
\caption{: \textsf{Edgeless-GNN}}
\label{mainalgorithm}
 \begin{algorithmic}[1]
  \renewcommand{\algorithmicrequire}{\textbf{Input:}}
  \renewcommand{\algorithmicensure}{\textbf{Output:}}
  \REQUIRE $G$, $\mathcal{X}$, $\mathcal{X}'$, $\theta$, $\alpha, num\_epochs$
  \ENSURE  $\mathbf{Z}'$
  \STATE \textbf{Initialization: } $G_{p} \leftarrow f_p(\emptyset, \mathcal{X})$; \\ $G_{p}^{\textnormal{all}} \leftarrow f_p(G_{p}, \mathcal{X}')$;\\$\theta \leftarrow \textnormal{random initialization}$
  \STATE /* Training phase */
  \FOR {$i=1,\cdots, num\_epochs$}
  \STATE $\mathbf{Z} \leftarrow \textsf{GNN}_{\theta}(G_{p}, \mathcal{X})$
  \STATE $\mathcal{L}({\bf Z}, \mathcal{E}) \leftarrow  \mathcal{L}_{0}({\bf Z}, \mathcal{E}) + \alpha \mathcal{L}_{\textnormal{2nd}}({\bf Z}, \mathcal{E})$
  \STATE Update $\theta$ by taking one step of gradient descent
  \ENDFOR
  \STATE /* Inference phase */
  \STATE $\mathcal{X}^{\textnormal{all}} \leftarrow \mathcal{X} \cup \mathcal{X}'$
  \STATE $\mathbf{Z}' \leftarrow \textsf{GNN}_{\theta}(G_{p}^{\textnormal{all}}, \mathcal{X}^{\textnormal{all}})$
  \RETURN $\mathbf{Z}'$
 \end{algorithmic}
\end{algorithm}

In this subsection, as another core component of \textsf{Edgeless-GNN}, we elaborate on our new loss function that is used during the training phase of the GNN model. By training the parameters of the GNN model using our loss function, we expect that the model learns how to bridge the gap between the attribute-based computation graph and the observed topological structure. To this end, we first generate multiple samples of a node quadruplet $(v_i,v_j,v_n,v_t)$, where $(v_i,v_j) \in \mathcal{E}$, $(v_i,v_n) \notin \mathcal{E}$, and $v_t$ is a two-hop neighbor of $v_i$. The sampled quadruplets are then fed into the loss function in the training loop along with the calculated embedding $\mathbf{Z} = \textsf{GNN}_{\theta}(G_{p}, \mathcal{X})$ for all nodes in $\mathcal{V}$ (refer to lines 4--5 in Algorithm~\ref{mainalgorithm}). 

In \textsf{Edgeless-GNN}, we design a topology-aware loss function for unsupervised learning:
\begin{equation}
\label{mainlossfunction}
    \mathcal{L}({\bf Z}, \mathcal{E}) =  \mathcal{L}_{0}({\bf Z}, \mathcal{E}) + \alpha \mathcal{L}_{\textnormal{2nd}}({\bf Z}, \mathcal{E}),
\end{equation}
which is based on the {\em energy-based learning} in~\cite{lecun2006energy} that aims at training a model in the sense of minimizing the energy of node pairs. Here, $\alpha > 0$ is a hyperparameter that balances between the two loss terms in (\ref{mainlossfunction}). Our loss function enables us to exploit not only the first-order proximity but also the {\it second-order positive} relations in learning node representations from the network structure. That is, the topology-aware loss function is designed to judiciously encode positive, negative, and second-order relations between nodes.

The first term $\mathcal{L}_0$ in (\ref{mainlossfunction}) is defined as
\begin{equation}
\label{L0loss}
    \mathcal{L}_{0}({\bf Z}, \mathcal{E}) = E^{+}_{ij} + D_{in} E^{-}_{in},
\end{equation}
where $E^{+}_{ij}$ and $E^{-}_{in}$ denote generic energy functions of positive node pairs $(v_i,v_j)$ and negative node pairs $(v_i,v_n)$, respectively; and 
\begin{equation}
\label{Dloss}
    D_{in} = \exp{\left( \dfrac{\beta}{d_{sp}(v_{i}, v_{n})}\right)}.
\end{equation} 
Here, $d_{sp}(v_{i}, v_{n})$ is the shortest distance between $v_{i}$ and $v_{n}$, and $\beta > 0$ is a hyperparameter controlling negative node pairs. For long $d_{sp}(v_i,v_n)$, the term $E_{in}^{-}$ corresponding to the energy function of negative node pairs would contribute less to $\mathcal{L}_0$. In our study, as in~\cite{hao2020inductive}, we set 
\begin{align}
E_{ij}^{+} &= \phi(\textnormal{sim}(\mathbf{z}_{i}, \mathbf{z}_{j}))\\
E_{in}^{-} &= \phi(-\textnormal{sim}(\mathbf{z}_{i}, \mathbf{z}_{n})),
\end{align}
where $\textnormal{sim}(\cdot)$ is the cosine similarity and $\phi(x) = \gamma^{-1}\log(1+\exp(-\gamma x + b))$ for hyperparameters $\gamma > 0$ and $b \geq 0$. 

While (\ref{L0loss}) is a good representation of the pairwise ranking-based loss~\cite{bojchevski2017deep,dinyuanzhuDVNE,hao2020inductive} in effectively capturing the network structure, only the first-order proximity for positive node pairs is taken into account. Motivated by the fact that higher-order positive relations of node pairs are also proven to be useful to enhance the performance of network embedding methods~\cite{perozzi2014deepwalk,grover2016node2vec, wang2016structural,dinyuanzhuDVNE}, we introduce $\mathcal{L}_{\textnormal{2nd}}$ to incorporate the ranking of nodes with respect to the {\em second-order} proximity as follows:
\begin{equation}
\label{Lalphaloss}
    \mathcal{L}_{\textnormal{2nd}}({\bf Z}, \mathcal{E}) = J_{it}E_{it}^{+},
\end{equation}
where $J_{it}$ is the Jaccard similarity~\cite{DBLP:books/mg/SaltonG83}, which measures the degree of the second-order proximity of node pairs $(v_i,v_t)$. This is because not all two-hop neighbors have high second-order proximities, and only considering two-hop neighbors is a good trade-off between the performance and the computational overhead. 
\begin{remark}
The red box in the training phase of Fig.~\ref{Schematicoverview} illustrates the effect of each term in our topology-aware loss, with four nodes ($v_1, v_2, v_4, v_5$) and their corresponding representations (${\bf z}_1, {\bf z}_2, {\bf z}_4, {\bf z}_5$). The first term forces the representations of the positive node pair $v_1$ and $v_2$ ({\it i.e.}, ${\bf z}_1$ and ${\bf z}_2$) to be closer with each other and the representations of the negative node pair $v_1$ and $v_5$ ({\it i.e.}, ${\bf z}_1$ and ${\bf z}_5$) to be further apart. That is, attractive and repulsive forces are given to the positive and negative node pairs, respectively, on the representation space. Additionally, the second term acting on the representations of the second-order positive node pair $v_1$ and $v_4$ ({\it i.e.}, ${\bf z}_1$ and ${\bf z}_4$) also forces to be close with each other while pulling the two nodes depending on the degree of the second-order proximity..
\end{remark}

After the loss $\mathcal{L}({\bf Z}, \mathcal{E})$ is calculated, the model parameters $\theta$ are updated using gradient descent optimization (refer to line 6 in Algorithm \ref{mainalgorithm}). 

\subsection{Complexity Analysis} \label{computcomplex}
In this subsection, we analyze the computational complexity of our proposed \textsf{Edgeless-GNN} framework in which the $k$NNG is constructed as a proxy graph.

\begin{theorem} \label{complexitythm}
The computational complexity of the \textsf{Edgeless-GNN} framework is at most {\em linear} in $k$ and $|\mathcal{V}|$.
\end{theorem}

\begin{proof}
The feed-forward process of GNNs includes the computation of both AGGREGATE and UPDATE functions, and its computational complexity is given by $\mathcal{O}(2|\mathcal{E}|+|\mathcal{V}|)$ (refer to~\cite{wu2019survey} for more details). In our \textsf{Edgeless-GNN} framework, the number of edges in a computation graph depends on $f_p(\cdot, \cdot)$ when we choose $k$NNG as our implementation. For the best case where all edges are created by selecting $k$ neighbors of each node in the sense of minimizing the graph density, $k|\mathcal{V}|/2$ edges are generated, thus yielding the computational complexity of $\Omega((k+1)|\mathcal{V}|)$. For the worst case, corresponding to the mutually exclusive selection of edges, $k|\mathcal{V}|$ edges are generated; the complexity is thus bounded by $\mathcal{O}((2k+1)|\mathcal{V}|)$. Hence, the computational complexity of our \textsf{Edgeless-GNN} is finally given by $\mathcal{O}(k|\mathcal{V}|)$, indicating a {\em linear} complexity in $k$ and $|\mathcal{V}|$, which completes the proof of Theorem~\ref{complexitythm}.
\end{proof}


\begin{table}
  \caption{Summary of statistics of five datasets, where NN, NE, NA, and NC denote the number of nodes, the number of edges, the number of node attributes, the number of classes, respectively.}
  \label{Tabledataset}
  \scalebox{1}{\begin{tabularx}{\columnwidth}{lYYYY}
        \toprule
        \textbf{Dataset} & \textbf{NN} & \textbf{NE} & \textbf{NA}  & \textbf{NC} \\
        \midrule
        Cora  & 2,485 & 5,069 & 1,433  & 7 \\
        Citeseer & 2,120 & 3,679 & 3,703 & 6\\
        Wiki & 2,357 & 12,714 & 4,973 &  19\\
        Pubmed & 19,717 & 44,324 & 500 & 3\\
        Coauthor-CS & 18,333 & 81,894 & 6,805 & 15\\
        \bottomrule
\end{tabularx}}
\end{table}

\section{Experimental Evaluation} \label{experimentalevaluation}
In this section, we first describe real-world datasets used in the evaluation. We also present downstream tasks with their performance metrics. After describing our experimental settings, we comprehensively evaluate the performance of our \textsf{Edgeless-GNN} framework and five benchmark methods including two variants of GNN models.

\subsection{Datasets}
Five real-world attributed network datasets, commonly adopted from the literature of {\em attributed} network embedding, are used to acquire the network structure $G$ and the set of node attributes, $\mathcal{X}^{\text{all}}$. For all experiments, we consider the largest connected component without isolated nodes to ensure that no edgeless nodes with no ground truth edges occur during the inference phase. The main statistics of each dataset are summarized in Table \ref{Tabledataset}. In the following, we describe important characteristics of the datasets.

\textbf{Cora}, \textbf{Citeseer}~\cite{sen2008citeseer}, and \textbf{Pubmed}~\cite{namata2012pubmed}. The three datasets are citation networks. Each node is a publication from various research topics, each of which represents the class label, and an edge exists if one publication cited another. The attribute matrix is the bag-of-word representation comprising a corpus of documents for Cora and Citeseer and is the representation weighted by term frequency-inverse document frequency (TF-IDF) for Pubmed. We use the version provided in~\cite{DBLP:conf/icml/YangCS16} for our experiments.

\textbf{Wiki}~\cite{yang2015tadw}. The Wiki dataset is a network of web pages where the nodes are web documents in Wikipedia. Edges are constructed when two web pages have hyperlinks. The attribute matrix is the representation weighted by TF-IDF.

\textbf{Coauthor-CS}~\cite{shchur2018coauthor}. The Coauthor-CS dataset is a co-authorship network where nodes are authors and are connected if they are co-authors of a paper. The attribute matrix includes keywords in each author's papers, and the class labels indicate the most active research field for each author.

\subsection{Downstream Tasks With Performance Metrics}
To empirically validate the performance of the proposed framework over the above benchmark methods in an inductive setting, we consider three downstream ML tasks and assess the performance via five metrics. We note that all metrics are in a range of $[0,1]$, and higher values represent better performance. The performance of each ML task is evaluated on {\em edgeless nodes} since we focus on the inductive setting along with inductive representation learning of edgeless nodes.

\textbf{Link prediction}~\cite{kipf2016vgae} aims to predict edges that are likely to be existent. In our study, we predict edges to which {\em edgeless nodes} are incident by obtaining a reconstructed graph $\hat{G}^\text{all}$ via calculated embeddings, i.e., $\hat{G}^\text{all}=\sigma({\bf Z}^\text{all}{\bf Z}^{\text{all}\top})$, where $\top$ denotes the transpose of a matrix. That is, we focus on predicting the set of two types of edges including the edges connecting two nodes in $\mathcal{V}'$ and the edges connecting one node in $\mathcal{V}$ and another node in $\mathcal{V}'$. We adopt the average precision (AP) and area under curve (AUC) scores for this task.

\textbf{Node classification}~\cite{perozzi2014deepwalk,DBLP:conf/nips/HamiltonYL17} aims to classify new nodes into their ground truth classes. We train a logistic regression classifier using a portion of node embeddings in a supervised manner. We adopt the macro-$F_1$ and micro-$F_1$ scores for this task.

\textbf{Community detection}~\cite{pan2018arge} aims to unsupervisedly find the set of communities. We apply $k$-means clustering, one of standard clustering techniques, to the embeddings for all nodes in $\mathcal{V}^{\text{all}}$ and then assign a community label to each node to predict ground-truth communities of new nodes. We adopt the normalized mutual information (NMI) for this task.

\subsection{Benchmark Methods}
\label{section5.2benchmark}
In this subsection, we present two state-of-the-art methods for inductive edgeless network embedding and one baseline method for comparison.

\textbf{DEAL}~\cite{hao2020inductive}. This state-of-the-art approach aims to solve the inductive link prediction problem. To this end, the embeddings generated by two encoders, namely an MLP encoder using node attributes and a linear encoder using one-hot node representations are aligned to learn the connections between the node attributes and the network structure so that the encoder generates a link prediction score for edgeless nodes.

\textbf{Graph2Gauss (G2G)}~\cite{bojchevski2017deep}. As another state-of-the-art method, the G2G model trains an MLP encoder that represents each node as a Gaussian distribution to capture the uncertainties of embeddings. The model also investigates its application to the problem of inductive link prediction.

\textbf{Inference only with node attributes (Att-Only)}. As a baseline, we include the case where the attribute matrix is only used for inference. In other words, the attribute matrix itself is taken into account as node embeddings.



\subsection{Experimental Settings}

We first describe the settings of neural networks. In our study, we choose GraphSAGE~\cite{DBLP:conf/nips/HamiltonYL17}, a widely used GNN architecture, as the base model for our framework. We train our GNN model via the Adam optimizer~\cite{kingma2015adam} with a learning rate of 0.0005 and a weight decay rate of 0.0005. The dimension of the embedding space is set to 64. All models were implemented in Python 3.7.7, PyTorch 1.5.1, and PyTorch Geometric 1.6~\cite{fey2019pyg}. The experiments were run on a machine with Intel Core i7-9700K 3.60 GHz CPU with 32GB RAM and one NVIDIA GeForce RTX 2080 graphics card.

In the $k$NNG construction, we set $k=3$ for all experiments. This is because 1) our experimental findings reveal that the performance is insensitive to the value of $k$ (refer to Table \ref{kexperiment} for experimental results in terms of link prediction) and 2) the value of $k$ needs to be set as small as possible since the computational complexity of \textsf{Edgeless-GNN} scales linearly with $k$ (refer to Section \ref{computcomplex}).

From each dataset, we randomly split the set of nodes into training/validation/test sets with a ratio of 85/5/10\%. In order to simulate our inductive setting, validation and test sets accounting for 15\% of nodes are treated as new edgeless nodes and assumed not known during training. We use the validation set to tune the hyperparameters for each downstream task and determine the number of training iterations. That is, we use the values of hyperparameters (e.g., $\alpha$ and $\beta$ presented in our loss function) optimally found according to different datasets and downstream ML tasks. During tuning, we use the following hyperparameter values within designated ranges: $\alpha \in \{2,3,4\}$, $b \in \{0,1\}$, $\beta \in \{1,2\}$, and $\gamma \in \{2,3,4\}$. For each evaluation, we run experiments over 10 different splits of training/validation/test sets to compute the average score.

\begin{table*}[h!]
\centering
\caption{Performance comparison among \textsf{Edgeless-GNN} and benchmark methods in terms of five performance metrics (average $\pm$ standard deviation). Here, the best and second best performers are highlighted by bold and underline, respectively.}
\label{Q3table}
\scalebox{1}{\begin{tabular}{ccccccc}
\toprule 
\multicolumn{1}{c}{} & {} & \multicolumn{2}{c}{Link prediction} & \multicolumn{2}{c}{Node classification} & Community detection\\
 \cmidrule{3-7} 
  Dataset & Method & AP & AUC & Macro-$F_1$ & Micro-$F_1$ & NMI \\ 
\midrule
\multirow{4}{*}{\rotatebox{0}{Cora}}  & \textsf{Edgeless-GNN} & \textbf{0.8930} $\pm$ 0.0140& \textbf{0.8905} $\pm$ 0.0127 & \textbf{0.6783} $\pm$ 0.0140 & \textbf{0.7177} $\pm$ 0.0343 & \textbf{0.5109} $\pm$ 0.0212\\ 
  & DEAL & \underline{0.8550} $\pm$ 0.0134 & \underline{0.8585} $\pm$ 0.0105 & \underline{0.6410} $\pm$ 0.0332 & \underline{0.6903} $\pm$ 0.0270& \underline{0.4321} $\pm$ 0.0193\\ 
  & G2G & 0.7966 $\pm$ 0.0470 & 0.8113 $\pm$ 0.0205 & 0.5983 $\pm$ 0.0165 & 0.6346 $\pm$ 0.0328 & 0.4089 $\pm$ 0.0354\\
  & Att-Only& 0.7546 $\pm$ 0.0126 & 0.7584 $\pm$ 0.0129 & 0.4923 $\pm$ 0.0347 & 0.5681 $\pm$ 0.0287 & 0.2213 $\pm$ 0.0602\\ 
\midrule
\multirow{4}{*}{\rotatebox{0}{Citeseer}} & \textsf{Edgeless-GNN} & \textbf{0.9385} $\pm$ 0.0062 & \textbf{0.9313} $\pm$ 0.0072 & \textbf{0.5832} $\pm$ 0.0378 & \underline{0.6697} $\pm$ 0.0299 & \textbf{0.4497} $\pm$ 0.0506\\ 
  & DEAL & \underline{0.9128} $\pm$ 0.0063 & \underline{0.9059} $\pm$ 0.0074 & \underline{0.5733} $\pm$ 0.0445 & \textbf{0.6701} $\pm$ 0.0350 & 0.3984 $\pm$ 0.0362 \\ 
  & G2G & 0.8545 $\pm$ 0.0721 & 0.8619 $\pm$ 0.0149 & 0.5424 $\pm$ 0.0110 & 0.6417 $\pm$ 0.0415 & \underline{0.4131} $\pm$ 0.0363\\ 
  & Att-Only & 0.8535 $\pm$ 0.0110 & 0.8448 $\pm$ 0.0112 & 0.5186 $\pm$ 0.0399 & 0.6293 $\pm$ 0.0402 & 0.2884 $\pm$ 0.0407\\
\midrule
\multirow{4}{*}{\rotatebox{0}{Wiki}}  & \textsf{Edgeless-GNN} & \textbf{0.7241} $\pm$ 0.0157 & \textbf{0.6842} $\pm$ 0.0211 & \textbf{0.5340} $\pm$ 0.0316 & \textbf{0.6596} $\pm$ 0.0273 & \textbf{0.6061} $\pm$ 0.0355\\ 
  & DEAL & \underline{0.6724} $\pm$ 0.0220 & \underline{0.6622} $\pm$ 0.0217 & 0.2065 $\pm$ 0.0256 & 0.4017 $\pm$ 0.0285 & \underline{0.5561} $\pm$ 0.0264 \\ 
  & G2G & 0.6026 $\pm$ 0.0265 & 0.6088 $\pm$ 0.0177 & \underline{0.4124} $\pm$ 0.0596 & \underline{0.5859} $\pm$ 0.0308 & 0.5213 $\pm$ 0.0488 \\ 
  & Att-Only & 0.6206 $\pm$ 0.0078 & 0.5725 $\pm$ 0.0095 & 0.2802 $\pm$ 0.0375 & 0.4557 $\pm$ 0.0418 & 0.4057 $\pm$ 0.0432\\
\midrule
\multirow{4}{*}{\rotatebox{0}{Pubmed}}  & \textsf{Edgeless-GNN} & \textbf{0.9413} $\pm$ 0.0033 & \textbf{0.9426} $\pm$ 0.0026& \textbf{0.8307} $\pm$ 0.0104 & \textbf{0.8306} $\pm$ 0.0102 & 0.3051 $\pm$ 0.0242\\ 
  & DEAL & 0.8974 $\pm$ 0.0025 & \underline{0.9119} $\pm$ 0.0024 & 0.8247 $\pm$ 0.0048 & 0.8264 $\pm$ 0.0040 & \textbf{0.3285} $\pm$ 0.0119\\ 
  & G2G & 0.8535 $\pm$ 0.0085 & 0.8756 $\pm$ 0.0062 & \underline{0.8278} $\pm$ 0.0106 & \underline{0.8304} $\pm$ 0.0094 & 0.3101 $\pm$ 0.0285\\ 
  & Att-Only & \underline{0.8977} $\pm$ 0.0047 & 0.8878 $\pm$ 0.0054 & 0.8164 $\pm$ 0.0077 & 0.8155 $\pm$ 0.0068 & \underline{0.3217} $\pm$ 0.0091\\
\midrule
\multirow{4}{*}{\shortstack{Coauthor-\\CS}}  & \textsf{Edgeless-GNN} & \textbf{0.9531} $\pm$ 0.0017 & \textbf{0.9503} $\pm$ 0.0017 & \textbf{0.8932} $\pm$ 0.0122 & \textbf{0.9218} $\pm$ 0.0058 & \textbf{0.7981} $\pm$ 0.0141 \\ 
  & DEAL & \underline{0.9342} $\pm$ 0.0020 & \underline{0.9319} $\pm$ 0.0021 & \underline{0.8689} $\pm$ 0.0106 & \underline{0.9124} $\pm$ 0.0065 & 0.6631 $\pm$ 0.0113\\ 
  & G2G & 0.8234 $\pm$ 0.0097 & 0.8516 $\pm$ 0.0051 & 0.8120 $\pm$ 0.0140 & 0.8738 $\pm$ 0.0078 & \underline{0.6857} $\pm$ 0.0083\\ 
  & Att-Only & 0.9034 $\pm$ 0.0020 & 0.9096 $\pm$ 0.0018 & 0.8301 $\pm$ 0.0214 & 0.8924 $\pm$ 0.0111 & 0.5662 $\pm$ 0.0150\\ 
\bottomrule
\end{tabular}}
\end{table*}

\subsection{Experimental Results}
Our empirical study is designed to answer the following six key research questions.
\begin{itemize}
    \item \textit{RQ1.} How much does the \textsf{Edgeless-GNN} framework improve the performance of various downstream tasks over state-of-the-art methods of inductive edgeless network embedding?
    \item \textit{RQ2.} How much does the \textsf{Edgeless-GNN} framework improve the performance of various downstream tasks over variants of existing GNN models for edgeless nodes?
    \item \textit{RQ3.} How do model hyperparameters affect the performance of the \textsf{Edgeless-GNN} framework?
    \item \textit{RQ4.} How do underlying GNN models affect the performance of the \textsf{Edgeless-GNN} framework?
    \item \textit{RQ5.} How robust is our \textsf{Edgeless-GNN} framework to the noise of node attributes?
    \item \textit{RQ6.} How scalable is our \textsf{Edgeless-GNN} framework with vital parameters including the graph size?
\end{itemize}

\begin{table*}[t!]
\centering
\caption{Performance comparison among \textsf{Edgeless-GNN} and two variants of GNN models in terms of five performance metrics (average $\pm$ standard deviation). Here, the best performers are highlighted by bold.}
\label{Abltable}
\scalebox{1}{\begin{tabular}{ccccccc}
\toprule 
\multicolumn{1}{c}{} & {} & \multicolumn{2}{c}{Link prediction} & \multicolumn{2}{c}{Node classification} & Community detection\\
 \cmidrule{3-7} 
  Dataset & Method & AP & AUC & Macro-$F_1$ & Micro-$F_1$ & NMI \\ 
\midrule
\multirow{3}{*}{Cora} & \textsf{Edgeless-GNN} & \textbf{0.8930} $\pm$ 0.0140& \textbf{0.8905} $\pm$ 0.0127 & \textbf{0.6783} $\pm$ 0.0140 & \textbf{0.7177} $\pm$ 0.0343 & \textbf{0.5109} $\pm$ 0.0212\\ 
  & SAGE-$k$NNG1 & 0.5961 $\pm$ 0.0193 & 0.5713 $\pm$ 0.0203 & 0.1990 $\pm$ 0.0481 & 0.3008 $\pm$ 0.0466 & 0.0615 $\pm$ 0.0228\\
  & SAGE-$k$NNG2 & 0.5889 $\pm$ 0.0208 & 0.5672 $\pm$ 0.0186 & 0.2100 $\pm$ 0.0167 & 0.3068 $\pm$ 0.0197 & 0.0569 $\pm$ 0.0206\\
\midrule
\multirow{3}{*}{Citeseer} & \textsf{Edgeless-GNN} & \textbf{0.9385} $\pm$ 0.0062 & \textbf{0.9313} $\pm$ 0.0072 & \textbf{0.5832} $\pm$ 0.0378 & \textbf{0.6697} $\pm$ 0.0299 & \textbf{0.4497} $\pm$ 0.0506\\ 
  & SAGE-$k$NNG1 & 0.5816 $\pm$ 0.0216 & 0.5588 $\pm$ 0.0236 & 0.1926 $\pm$ 0.0443 & 0.2995 $\pm$ 0.0264 & 0.0585 $\pm$ 0.0178\\
  & SAGE-$k$NNG2 & 0.6024 $\pm$ 0.0250 & 0.6035 $\pm$ 0.0317 & 0.2004 $\pm$ 0.0316 & 0.3023 $\pm$ 0.0403 & 0.0458 $\pm$ 0.0116\\
\midrule
\multirow{3}{*}{Wiki}  & \textsf{Edgeless-GNN} & \textbf{0.7241} $\pm$ 0.0157 & \textbf{0.6842} $\pm$ 0.0211 & \textbf{0.5340} $\pm$ 0.0316 & \textbf{0.6596} $\pm$ 0.0273 & \textbf{0.6061} $\pm$ 0.0355\\ 
  & SAGE-$k$NNG1 & 0.5730 $\pm$ 0.0066 & 0.5329 $\pm$ 0.0150 & 0.0258 $\pm$ 0.0067 & 0.1702 $\pm$ 0.0232 & 0.1589 $\pm$ 0.0216\\
  & SAGE-$k$NNG2 & 0.5648 $\pm$ 0.0134 & 0.5300 $\pm$ 0.0090 & 0.0327 $\pm$ 0.0061 & 0.1714 $\pm$ 0.0290 & 0.1469 $\pm$ 0.0270\\
\midrule
\multirow{3}{*}{Pubmed} & \textsf{Edgeless-GNN} & \textbf{0.9413} $\pm$ 0.0033 & \textbf{0.9426} $\pm$ 0.0026& \textbf{0.8307} $\pm$ 0.0104 & \textbf{0.8306} $\pm$ 0.0102 & \textbf{0.3051} $\pm$ 0.0242\\ 
  & SAGE-$k$NNG1 & 0.6077 $\pm$ 0.0104 & 0.6074 $\pm$ 0.0112 & 0.3867 $\pm$ 0.0165 & 0.4560 $\pm$ 0.0121 & 0.0196 $\pm$ 0.0145\\
  & SAGE-$k$NNG2 & 0.6191 $\pm$ 0.0086 & 0.6210 $\pm$ 0.0111 & 0.3830 $\pm$ 0.0145 & 0.4498 $\pm$ 0.0212 & 0.0263 $\pm$ 0.0238\\
\midrule
\multirow{3}{*}{\shortstack{Coauthor-\\CS}}  & \textsf{Edgeless-GNN} & \textbf{0.9531} $\pm$ 0.0017 & \textbf{0.9503} $\pm$ 0.0017 & \textbf{0.8932} $\pm$ 0.0122 & \textbf{0.9218} $\pm$ 0.0058 & \textbf{0.7981} $\pm$ 0.0141 \\ 
  & SAGE-$k$NNG1 & 0.7074 $\pm$ 0.0035 & 0.6872 $\pm$ 0.0031 & 0.0356 $\pm$ 0.0034 & 0.2363 $\pm$ 0.0080 & 0.0674 $\pm$ 0.0233\\
  & SAGE-$k$NNG2 & 0.7149 $\pm$ 0.0074 & 0.6952 $\pm$ 0.0055 & 0.0704 $\pm$ 0.0074 & 0.2513 $\pm$ 0.0123 & 0.0436 $\pm$ 0.0114\\
\bottomrule
\end{tabular}}
\end{table*}

\subsubsection{RQ1. Comparison with Benchmark Methods}
Table \ref{Q3table} shows the performance comparison between our \textsf{Edgeless-GNN} framework and three benchmark methods for inductive network embedding, including DEAL~\cite{hao2020inductive}, G2G~\cite{bojchevski2017deep}, and Att-Only, with respect to five performance metrics using all five real-world datasets. From the experimental results, we observe the following:
\begin{enumerate}
    \item Our \textsf{Edgeless-GNN} framework consistently outperforms state-of-the-art methods except for only one case each in node classification and community detection. Note that, for the Citeseer dataset, the micro-$F_1$ score achieved by \textsf{Edgeless-GNN} is almost the same as that of DEAL, exhibiting the performance gap of only 0.04\%; this indicates that the class labels have strong discriminative information and thus additional gains from $k$NNG construction are relatively low.
    \item The performance gap between our \textsf{Edgeless-GNN} and the second best method is the largest when we perform node classification using the Wiki dataset; the maximum improvement rates of $29.5\%$ and $12.5\%$ are achieved in terms of macro-$F_1$ and micro-$F_1$ scores, respectively. This coincides with the argument in~\cite{zhu2020h2gcn}, where GraphSAGE, the underlying GNN model applied to our \textsf{Edgeless-GNN}, is shown to be robust to networks even with low homophily ratios such as Wiki.
    \item The G2G method~\cite{bojchevski2017deep} tends to exhibit poor performance, often being even inferior to Att-Only, on the link prediction task for the Pubmed and Coauthor-CS datasets. This is because the Kullback-Leibler (KL) divergence, used in the loss of G2G to measure the similarity of node pairs, is asymmetric and does not well preserve the transitivity of proximity in undirected graphs as discussed in~\cite{dinyuanzhuDVNE}, thus leading to distortion in the embedding space to some extent.
    \item For a given dataset, the highest gain of \textsf{Edgeless-GNN} over benchmark methods tends to be achieved in either node classification or community detection. This comes from the fact that DEAL~\cite{hao2020inductive}, the second best performer, was originally designed for the link prediction task.
    \item The performance gain of \textsf{Edgeless-GNN} over benchmark methods is higher mostly in the community detection task than that in the node classification task. As \textsf{Edgeless-GNN} utilizes the $k$NNG which already connects nodes with similar attributes, it is generally advantageous to community detection when the network community structure follows node attributes. However, for the Pubmed dataset, we observe that the overall scores for community detection are low but also Att-Only performs quite well, which may indicate that that the attribute information does not agree with the underlying network structure. Therefore, such intrinsic characteristics of Pubmed are interpreted as a factor that impairs the performance as \textsf{Edgeless-GNN} attempts to capture the network structure by the learned parameters. 
\end{enumerate}

\begin{table*}[t]
  \caption{Performance of our \textsf{Edgeless-GNN} framework for the Cora and Citeseer datasets according to different values of $k$.} \label{kexptable}
  \centering
  \begin{subtable}{1.8\columnwidth}\centering
  {
  \begin{tabularx}{\columnwidth}{lYYYYY}
        \toprule
        $k$ & 2 & 3 & 4  & 5 & 6\\
        \midrule
        AP  & 0.8852 $\pm$ 0.0096 & 0.8821 $\pm$ 0.0105 & 0.8763 $\pm$ 0.0103  & 0.8886 $\pm$ 0.0095 & 0.8830 $\pm$ 0.0127\\
        AUC  & 0.8831 $\pm$ 0.0089 & 0.8775 $\pm$ 0.0102 & 0.8752 $\pm$ 0.0081 & 0.8851 $\pm$ 0.0086 & 0.8823 $\pm$ 0.0081 \\
        Macro-$F_1$  & 0.6213 $\pm$ 0.0338 & 0.6520 $\pm$ 0.0336 & 0.6279 $\pm$ 0.0374 & 0.6197 $\pm$ 0.0352 & 0.6356 $\pm$ 0.0405 \\
        Micro-$F_1$  & 0.6629 $\pm$ 0.0308 & 0.6940 $\pm$ 0.0275 & 0.6738 $\pm$ 0.0362  & 0.6737 $\pm$ 0.0290 & 0.6899 $\pm$ 0.0482\\
        NMI  & 0.4709 $\pm$ 0.0450 & 0.5036 $\pm$ 0.0316 & 0.5086 $\pm$ 0.0383 & 0.4959 $\pm$ 0.0225 & 0.5053 $\pm$ 0.0265 \\
        \bottomrule
\end{tabularx}
\caption{Effect of $k$ on Cora.}\label{kexperiment}
}
\end{subtable}
\centering
\begin{subtable}{1.8\columnwidth}\centering
  {
  \begin{tabularx}{\columnwidth}{lYYYYY}
        \toprule
        $k$ & 2 & 3 & 4  & 5 & 6\\
        \midrule
        AP  & 0.9376  $\pm$ 0.0077 & 0.9385  $\pm$ 0.0095 & 0.9325 $\pm$ 0.0103  & 0.9365 $\pm$ 0.0082 & 0.9385 $\pm$ 0.0063 \\
        AUC  & 0.9279 $\pm$ 0.0072 & 0.9313 $\pm$ 0.0165 & 0.9282  $\pm$ 0.0095 & 0.9284  $\pm$ 0.0081 & 0.9304 $\pm$ 0.0078\\
        Macro-$F_1$  & 0.5752 $\pm$ 0.0398 & 0.5698 $\pm$ 0.0463 & 0.5589 $\pm$ 0.0564  & 0.5508  $\pm$ 0.0494 & 0.5767 $\pm$ 0.0456 \\
        Micro-$F_1$  & 0.6545 $\pm$ 0.0250 & 0.6511 $\pm$ 0.0487 & 0.6672  $\pm$ 0.0388 & 0.6620 $\pm$ 0.0377  & 0.6507 $\pm$ 0.0407\\
        NMI  & 0.4184 $\pm$ 0.0326 & 0.4126 $\pm$ 0.0377 & 0.4336 $\pm$ 0.0351  & 0.4589 $\pm$ 0.0453   & 0.4253 $\pm$ 0.0269\\
        \bottomrule
\end{tabularx}
\caption{Effect of $k$ on Citeseer.}\label{kexperimentciteseer}
}
\end{subtable}
\end{table*}

\subsubsection{RQ2. Comparison with Variants of GNN Models for Edgeless Nodes}

For performance comparison, we modify existing GNN models so that they are suitable for our settings with edgeless nodes. To this end, we present two variants built upon GraphSAGE as follows.

\textbf{SAGE-$k$NNG1}. This modified version of \textsf{Edgeless-GNN} adopts the loss function and the GNN architecture in GraphSAGE~\cite{DBLP:conf/nips/HamiltonYL17}, while using the underlying network $G$ as a computation graph during the training phase. However, during the inference phase, due to the absence of connectivity information on edgeless nodes, we generate additional edges on the existing $G$ via $k$NNG construction based on the similarity of node attributes $\mathcal{X}^{\textnormal{all}}$ for each edgeless node in order to perform message passing along with the trained GraphSAGE model. 

\textbf{SAGE-$k$NNG2}. Another variant adopts the loss function and the GNN architecture from GraphSAGE during the training phase. However, SAGE-$k$NNG2 utilizes the $k$NNG generated from node attributes for both training and inference, following our approach in \textsf{Edgeless-GNN}.

Table~\ref{Abltable} shows the performance comparison among \textsf{Edgeless-GNN} and two variants of GNN models with respect to five performance metrics using all datasets. From Tables~\ref{Q3table} and~\ref{Abltable}, we observe that the two $k$NNG-aided baseline methods, SAGE-$k$NNG1 and SAGE-$k$NNG2, exhibit quite unsatisfactory performance and are even far inferior to Att-only ({\em i.e.}, a na\"{i}ve baseline employing node attributes only). This indicates that the proxy graph construction alone is not sufficient to bring satisfactory performance and the thorough design of a loss function plays an important role. In contrast to GraphSAGE~\cite{DBLP:conf/nips/HamiltonYL17}, our newly designed loss function takes into account not only the shortest distance between negative node pairs but also the second-order positive relations of node pairs. Exploitation of the high-order proximity of node pairs is shown to be significant in boosting the performance of downstream tasks.

\begin{figure}[t]
\centering
\begin{subfigure}{0.8\columnwidth}
\centering
\includegraphics[width=\textwidth]{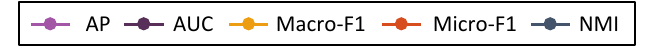}
\end{subfigure}
\begin{subfigure}{\columnwidth}
\centering
\includegraphics[width=\textwidth]{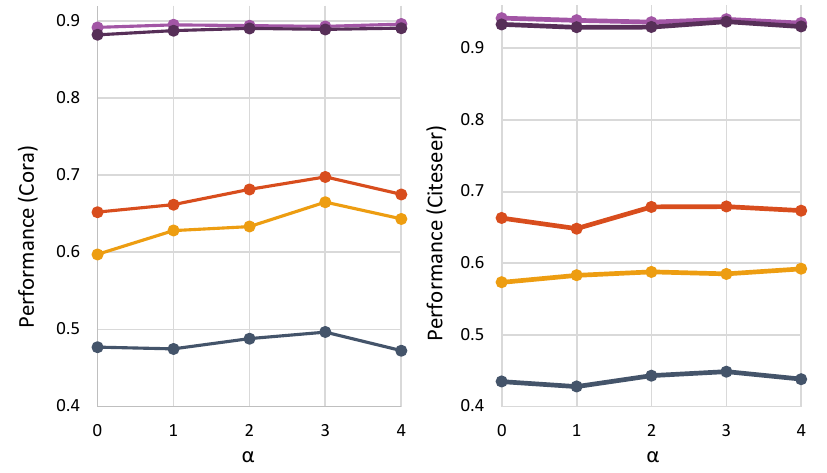}
\caption{Effect of $\alpha$ on Cora and Citeseer.}\label{Q2plot1alpha}
\end{subfigure}
\begin{subfigure}{\columnwidth}
\centering
\includegraphics[width=\textwidth]{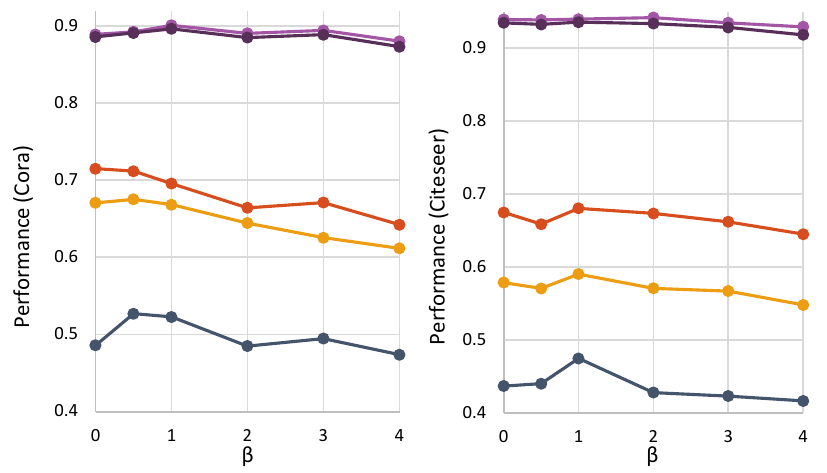}
\caption{Effect of $\beta$ on Cora and Citeseer.}\label{Q2plot1beta}
\end{subfigure}

\caption{Performance of downstream ML tasks according to different values of hyperpaprameters $\alpha$ and $\beta$ in our \textsf{Edgeless-GNN} framework validated for the Cora and Citeseer datasets.}
\label{Q2plot1}
\end{figure}

\subsubsection{RQ3. Hyperparameter Sensitivity}

We investigate the impact of hyperparameters $\alpha$, $\beta$, and $k$ on the performance of \textsf{Edgeless-GNN}. In our framework, $\alpha$ controls the strength of the second-order proximity loss term in (\ref{mainlossfunction}), $\beta$ controls the effect of negative node pairs in (\ref{Dloss}), and $k$ determines the number of edges in our $k$NNG implementation. For brevity, we report experimental results using the Cora and Citeseer datasets. Note that the results from other datasets showed a tendency similar to those reported in Section 5.4.3.\footnote{We refer to https://github.com/jordan7186/Edgeless-GNN-external.}

\begin{figure}[t]
\centering
\includegraphics[width=\columnwidth]{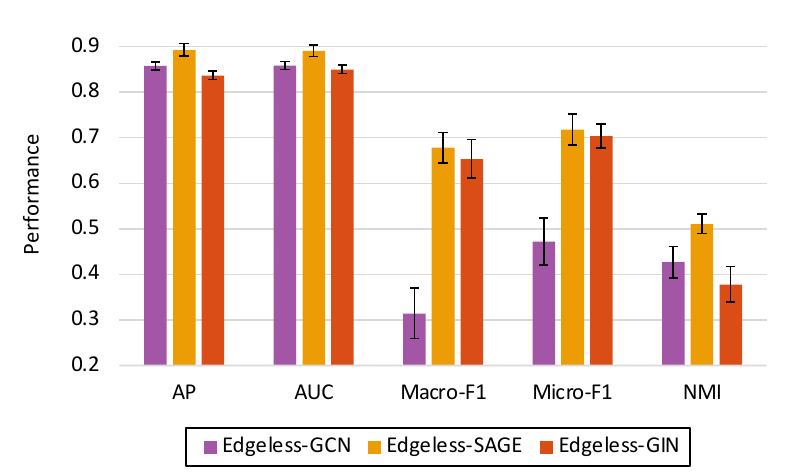}
\caption{Performance of downstream ML tasks according to different GNN models in our \textsf{Edgeless-GNN} framework for the Cora dataset.}
\label{Q1plot}
\end{figure}

In Fig.~\ref{Q2plot1alpha}, we plot the performance of downstream tasks according to different values of $\alpha$ while fixing the values of $\beta$ and $k$ to 1 and 3, respectively. For the Cora dataset, the maximum tends to be achieved at $\alpha=3$ regardless of downstream tasks. When $\alpha$ exceeds this value, the effect of the second-order proximity becomes over-amplified, thus leading to distorted embeddings. We observe a similar trend for the Citeseer dataset; however, the effect of over-amplification is less prominent, even achieving maximum performance at $\alpha = 4$ with respect to the Macro-$F_1$ score.
In Fig.~\ref{Q2plot1beta}, we plot the performance according to different values of $\beta$ while fixing the values of both $\alpha$ and $k$ to 3. Both plots in Fig.~\ref{Q2plot1beta} reveal that the performance tends to deteriorate in the case where $\beta>1$. This is because, in such a case, the exponent in (\ref{Dloss}) will explode, thus providing an {\it unbalanced repulsive force} between negative node pairs. In contrast, when $\beta=0$, the performance also gets reduced. Therefore, we conclude that setting $\beta$ to a low value near 1 achieves satisfactory performance. Overall, we also observe that the effect of $\alpha$ and $\beta$ is more prominent in node classification and community detection. 

Furthermore, Tables~\ref{kexperiment} and~\ref{kexperimentciteseer} show the performance according to different values of $k$ while fixing the values $\alpha$ and $\beta$ to 3 and 1, respectively. From Table~\ref{kexptable}, our findings reveal that the performance is insensitive to the value of $k$. Due to such robustness to $k$, one can choose a rather low value of $k$ ({\em e.g.}, $k=3$) to reduce the computational complexity in the feed-forward process of GNNs.

\begin{table*}[ht!]
\centering
\setlength\tabcolsep{6pt}
\scalebox{1}{
\begin{tabular}{cccccc}
\toprule
\multicolumn{1}{c}{} & {} & \multicolumn{4}{c}{\thead{The portion of missing node attributes (\%)}}\\
 \cmidrule{3-6} 
 Metric & Method & 20 & 40 & 60 & 80 \\ 
\midrule
\multirow{6}{*}{\rotatebox{90}{AP}}  & \textsf{Edgeless-GNN} & \textbf{0.8719} $\pm$ 0.0097 & \textbf{0.8517} $\pm$ 0.0124 & \textbf{0.8083} $\pm$ 0.0143 & \textbf{0.7875} $\pm$ 0.0157\\ 
  & DEAL & \underline{0.8328} $\pm$ 0.0092 & \underline{0.8050} $\pm$ 0.0056 & \underline{0.7755} $\pm$ 0.0122 & \underline{0.7514} $\pm$ 0.0117\\ 
  & G2G & 0.7503 $\pm$ 0.0131 & 0.7268 $\pm$ 0.0154 & 0.7047 $\pm$ 0.0154 & 0.6987 $\pm$ 0.0154\\ 
  & Att-Only& 0.7252 $\pm$ 0.0079 & 0.6837 $\pm$ 0.0106 & 0.6370 $\pm$ 0.0071 & 0.6038 $\pm$ 0.0109\\ 
  & SAGE-$k$NNG1 & 0.5712 $\pm$ 0.0195 & 0.5380 $\pm$ 0.0196 & 0.5287 $\pm$ 0.0158 & 0.5162 $\pm$ 0.0189\\ 
  & SAGE-$k$NNG2 & 0.5700 $\pm$ 0.0226 & 0.5349 $\pm$ 0.0197 & 0.5023 $\pm$ 0.0159 & 0.5160 $\pm$ 0.0139\\ 
\midrule
\multirow{6}{*}{\rotatebox{90}{AUC}}  & \textsf{Edgeless-GNN} & \textbf{0.8815} $\pm$ 0.0092 & \textbf{0.8642} $\pm$ 0.0160 & \textbf{0.8565} $\pm$ 0.0154 & \textbf{0.8408} $\pm$ 0.0103 \\ 
  & DEAL & \underline{0.7952} $\pm$ 0.0101 &  \underline{0.7918} $\pm$ 0.0081 & \underline{0.7900} $\pm$ 0.0129 & \underline{0.7921} $\pm$ 0.0131\\
  & G2G & 0.7945 $\pm$ 0.0131 & 0.7652 $\pm$ 0.0113 & 0.7508 $\pm$ 0.0186 & 0.7423 $\pm$ 0.0177\\ 
  & Att-Only& 0.7582 $\pm$ 0.0101 & 0.7365 $\pm$ 0.0073 & 0.7170 $\pm$ 0.0067 & 0.7001 $\pm$ 0.0100\\ 
  & SAGE-$k$NNG1 & 0.5589 $\pm$ 0.0203 & 0.5343 $\pm$ 0.0150 & 0.5222 $\pm$ 0.0183 & 0.5160 $\pm$ 0.0182\\ 
  & SAGE-$k$NNG2 & 0.5571 $\pm$ 0.0147 & 0.5378 $\pm$ 0.0194 & 0.5114 $\pm$ 0.0218 & 0.5201 $\pm$ 0.0141\\ 
\midrule
\multirow{6}{*}{\rotatebox{90}{Macro-$F_1$}}  & \textsf{Edgeless-GNN} & \textbf{0.5779} $\pm$ 0.0332 & \textbf{0.5603} $\pm$ 0.0404 & \textbf{0.5002} $\pm$ 0.0253 & \textbf{0.4580} $\pm$ 0.0461\\ 
  & DEAL & \underline{0.5755} $\pm$ 0.0378 & \underline{0.5157} $\pm$ 0.0262 & \underline{0.4884} $\pm$ 0.0261 & \underline{0.4412} $\pm$ 0.0297\\ 
  & G2G & 0.5250 $\pm$ 0.0322 & 0.4808 $\pm$ 0.0326& 0.4389 $\pm$ 0.0560 & 0.4083 $\pm$ 0.0244\\ 
  & Att-Only & 0.4214 $\pm$ 0.0429 & 0.3571 $\pm$ 0.0411 & 0.3330 $\pm$ 0.0394 & 0.2863 $\pm$ 0.0402\\
  & SAGE-$k$NNG1 & 0.1881 $\pm$ 0.0284 & 0.1676 $\pm$ 0.0353 & 0.1233 $\pm$ 0.0247 & 0.0854 $\pm$ 0.0130\\ 
  & SAGE-$k$NNG2 & 0.1847 $\pm$ 0.0334 & 0.1499 $\pm$ 0.0328 & 0.1020 $\pm$ 0.0164 & 0.0776 $\pm$ 0.0114\\ 
\midrule
\multirow{6}{*}{\rotatebox{90}{Micro-$F_1$}}  & \textsf{Edgeless-GNN} & \textbf{0.6701} $\pm$ 0.0290 & \underline{0.6229} $\pm$ 0.0449 & \underline{0.6161} $\pm$ 0.0444 & \underline{0.6028} $\pm$ 0.0311\\ 
  & DEAL & \underline{0.6459} $\pm$ 0.0277 & \textbf{0.6447} $\pm$ 0.0261 & \textbf{0.6375} $\pm$ 0.0285 & \textbf{0.6451} $\pm$ 0.0271\\ 
  & G2G & 0.6120 $\pm$ 0.0394 & 0.5903 $\pm$ 0.0280 & 0.5455 $\pm$ 0.0413 & 0.5548 $\pm$ 0.0110\\ 
  & Att-Only& 0.5399 $\pm$ 0.0250 & 0.5064 $\pm$ 0.0376 & 0.4536 $\pm$ 0.0434 & 0.4262 $\pm$ 0.0217\\ 
  & SAGE-$k$NNG1 & 0.3052 $\pm$ 0.0208 & 0.2822 $\pm$ 0.0328 & 0.2842 $\pm$ 0.0272 & 0.2939 $\pm$ 0.0252\\ 
  & SAGE-$k$NNG2 & 0.2802 $\pm$ 0.0154 & 0.2967 $\pm$ 0.0106 & 0.2641 $\pm$ 0.0246 & 0.2838 $\pm$ 0.0235\\ 
\midrule
\multirow{6}{*}{\rotatebox{90}{\thead{NMI}}}  & \textsf{Edgeless-GNN} & \textbf{0.4381} $\pm$ 0.0276 & \textbf{0.3922} $\pm$ 0.0357 & \textbf{0.3087} $\pm$ 0.0300 & \textbf{0.2838} $\pm$ 0.0262\\ 
  & DEAL & \underline{0.3638} $\pm$ 0.0360 & \underline{0.3442} $\pm$ 0.0271 & \underline{0.2866} $\pm$ 0.0383 & \underline{0.2398} $\pm$ 0.0269\\ 
  & G2G & 0.3397 $\pm$ 0.0355 & 0.2915 $\pm$ 0.0174 & 0.2444 $\pm$ 0.0252 & 0.2100 $\pm$ 0.0316\\ 
  & Att-Only & 0.1575 $\pm$ 0.0390 & 0.1198 $\pm$ 0.0248 & 0.0845 $\pm$ 0.0275 & 0.0720 $\pm$ 0.0211\\ 
  & SAGE-$k$NNG1 & 0.0497 $\pm$ 0.0118 & 0.0436 $\pm$ 0.0122 & 0.0496 $\pm$ 0.0096 & 0.0398 $\pm$ 0.0103\\ 
  & SAGE-$k$NNG2 & 0.0525 $\pm$ 0.0140 & 0.0405 $\pm$ 0.0110 & 0.0351 $\pm$ 0.0121 & 0.0428 $\pm$ 0.0078\\ 
\bottomrule
\end{tabular}}
\caption{Performance comparison among \textsf{Edgeless-GNN} and five benchmark methods in terms of five performance metrics (average $\pm$ standard deviation) using the Cora dataset when some of node attributes are missing. Here, the best method for each case is highlighted by bold and underline, respectively.}
\label{Tablemissingatt}
\end{table*}

\subsubsection{RQ4. Comparative Study Among Various GNN Models} \label{basegnnmodelstudy}

In Fig.~\ref{Q1plot}, we show the performance of different downstream tasks using various GNN models in our \textsf{Edgeless-GNN} framework using the Cora dataset. Note that the results from other datasets showed a tendency similar to those reported in Section~\ref{basegnnmodelstudy}. In our experiments, we adopt the following three widely used GNN models from the literature~\cite{wu2019survey}: GCN~\cite{DBLP:conf/iclr/KipfW17}  (\textsf{Edgeless-GCN}), GraphSAGE~\cite{DBLP:conf/nips/HamiltonYL17} (\textsf{Edgeless-SAGE}), and GIN~\cite{xu2018powerful} (\textsf{Edgeless-GIN}).

\textsf{Edgeless-SAGE} consistently outperforms other models regardless of downstream tasks. Such a gain is possible due to the concatenation operation of the UPDATE function, which enables the GraphSAGE model to elaborately assign different weights to the aggregated messages from neighbors, unlike other models with the mean/sum aggregator~\cite{zhu2020h2gcn}. For other two models, one does not always dominate another. \textsf{Edgeless-GCN} performs better than \textsf{Edgeless-GIN} in terms of link prediction and community detection while an opposite trend occurs for node classification.

\subsubsection{RQ5. Robustness to Incomplete Node Attributes} \label{incompletenodeattributessection}
We now evaluate the performance in a more practical setting where some node attributes are missing. To this end, we create {\em incomplete attributed} networks by randomly masking $\{20,40,60,80\}\%$ of node attributes. We only show the results for the Cora dataset since the results from other datasets follow similar trends. The missing information is masked as zero, indicating the absence of information. 

The performance comparison between our \textsf{Edgeless-GNN} framework and the five benchmark methods, including two variants of GNN models, is presented in Table \ref{Tablemissingatt} with respect to five performance metrics using the Cora dataset. Our findings demonstrate that, while the performance tends to degrade with an increasing portion of missing node attributes for all the methods, our \textsf{Edgeless-GNN} mostly achieves superior performance compared to other methods in terms of all performance metrics even for the case where only a small portion of node attributes are observed. This implies that our framework is robust to the degree of observability of node attributes even if a proxy graph is constructed using the set of incomplete node attributes.

\subsubsection{RQ6. Empirical validation of the Complexity analysis} \label{complexityanalysis}
To empirically validate our complexity analysis shown in Theorem 4.1, we conduct experiments using the Pubmed dataset whose number of nodes is sufficiently large in flexibly altering the graph size, where a full-batch setting is assumed. Specifically, we have performed two experiments, which is designed to validate the dependency of the complexity with respect to the number of nodes and the value of $k$, respectively. For the first experiment, we sample subgraphs of various sizes from Pubmed, as the number of nodes is large enough to perform experiments at multiple scales. Specifically, we create subgraphs having \{1,000, 3,000, 5,000, 7,000, 9,000, 11,000\} nodes using the forest fire sampling~\cite{leskovec2006graphsampling} so as to examine the scalability while preserving the same structural properties. For the second experiment, we use a subgraph of a fixed size with 2,500 nodes while varying the parameter $k$ in $k$\textsf{NNG} from 2 to 8, which generates multiple proxy graphs with different number of edges.
 
\begin{figure}[t]
\centering
\includegraphics[width=\columnwidth]{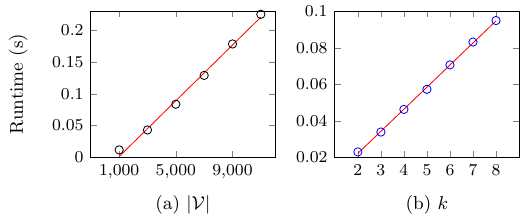}
\caption{The computational complexity of \textsf{Edgeless-GNN} with respect to $k$ and $|\mathcal{V}|$ for the Pubmed dataset, where the analytical results with proper biases are also plotted with red solid lines.}
\label{Q5plot}
\end{figure}

Fig.~\ref{Q5plot}a and Fig.~\ref{Q5plot}b show the measured runtime in seconds with respect to different $|\mathcal{V}|$'s and $k$'s, respectively. Asymptotic solid lines are also shown in the figure, showing a trend that is consistent with our experimental results. These results validate our analytical claim, i.e., a linear complexity scaling with respect to $k$ and $\mathcal{V}$.

\section{Concluding Remarks} \label{concludingremarks}
In this paper, we explored an important and challenging problem of how to exploit the power of GNNs in the context of embedding of topologically unseen nodes. To tackle this challenge, we introduced \textsf{Edgeless-GNN}, a novel GNN framework that unsupervisedly and inductively discovers vector representations of {\em edgeless nodes}. Specifically, we developed an approach to 1) constructing a new computation graph based on the similarity of node attributes to replace the original one used in GNNs and 2) then training our model by establishing our own topology-aware loss function that exploits not only the first-order proximity of node pairs but also second-order relations. Using five real-world datasets, we demonstrated that, for almost all cases, our \textsf{Edgeless-GNN} framework consistently outperforms state-of-the-art inductive network embedding methods for edgeless nodes, where the maximum gain of 29.5\% is achieved.

Potential avenues of future research include the design of a more robust and sophisticated \textsf{Edgeless-GNN} framework when a portion of node attributes are noisy and/or missing in attributed networks.

\ifCLASSOPTIONcompsoc
  \section*{Acknowledgments}
\else
  \section*{Acknowledgment}
\fi

This research was supported by the National Research Foundation of Korea (NRF) grant funded by the Korea government (MSIT) (No. 2021R1A2C3004345).

\ifCLASSOPTIONcaptionsoff
  \newpage
\fi



\bibliographystyle{IEEEtran}
\bibliography{main_tetc.bbl}
%



%

\end{document}